\documentclass[entropy,article,accept,pdftex,moreauthors]{Definitions/mdpi}

\firstpage{1} 
\pubvolume{1}
\issuenum{1}
\articlenumber{0}
\pubyear{2023}
\copyrightyear{2023}
\externaleditor{Academic Editor:}
\datereceived{} 
\daterevised{}
\dateaccepted{} 
\datepublished{} 
\hreflink{https://doi.org/} 

\Title{Cryptocurrencies Are Becoming Part of the World Global Financial Market}

\TitleCitation{Cryptocurrencies Are Becoming Part of the World Global Financial Market}

\Author{Marcin W\k atorek $^{1,}$*\orcidA{}, Jaros{\l}aw Kwapie\'n $^{2}$\orcidB{} and Stanis{\l}aw Dro\.zd\.z $^{1,2,}$*\orcidC{}}

\AuthorNames{Marcin Wątorek, Jarosław Kwapień and Stanisław Drożdż}

\AuthorCitation{Wątorek, M.; Kwapień, J.; Drożdż, S.}
\address{%
$^{1}$ \quad Faculty of Computer Science and Telecommunications, Cracow University of Technology, ul.~Warszawska 24, 31-155 Krak\'ow, 
Poland\\
$^{2}$ \quad Complex Systems Theory Department, Institute of Nuclear Physics, Polish Academy of Sciences, Radzikowskiego 152, 31-342 Kraków, Poland; jaroslaw.kwapien@ifj.edu.pl}

\corres{Correspondence: marcin.watorek@pk.edu.pl (M.W.); stanislaw.drozdz@ifj.edu.pl (S.D.)} 

\abstract{In this study the cross-correlations between the cryptocurrency market represented by the two most liquid and highest-capitalized cryptocurrencies: bitcoin and ethereum, on the one side, and the instruments representing the traditional financial markets: stock indices, Forex, commodities, on the other side, are measured in the period: January 2020--October 2022. {Our purpose is to address the question whether the cryptocurrency market still preserves its autonomy with respect to the traditional financial markets or it has already aligned with them in expense of its independence. We are motivated by the fact that some previous related studies gave mixed results.} By calculating the $q$-dependent detrended cross-correlation coefficient based on the high frequency 10 s data in the rolling window, the dependence on various time scales, different fluctuation magnitudes, and different market periods are examined. There is a strong indication that the dynamics of the bitcoin and ethereum price changes since the March 2020 Covid-19 panic is no longer independent. Instead, it is related to the dynamics of the traditional financial markets, which is especially evident now in 2022, when the bitcoin and ethereum coupling to the US tech stocks is observed during the market bear phase. It is also worth emphasizing that the cryptocurrencies have begun to react to the economic data such as the {Consumer Price Index} readings in a similar way as traditional instruments. Such a spontaneous coupling of the so far independent degrees of freedom can be interpreted as a kind of phase transition that resembles the collective phenomena typical for the complex systems. Our results indicate that the cryptocurrencies cannot be considered as a safe haven for the financial investments.}

\keyword{complexity; financial markets; cryptocurrencies; cross-correlations; multiscale; hedge} 

\begin{document}

\section{Introduction}
\label{sect::introduction}

From the physics point of view, the financial markets are considered as one of the most complex systems we observe in our world~\cite{kwapien2012}. Not only they are characterised by all the properties such systems can typically show, but there is also an important intelligent component involved that is decisively responsible for their enormous complexity. Among the well-known features of the financial markets is their flexibility in the transition between the disordered and ordered phases. Such a transition is the key feature associated with the market crashes but it is also often observed on the level of whole markets when some so-far independent markets start to have their dynamics substantially coupled (or \textit{vice-versa}).
 Exactly this kind of phenomenon has recently been experienced by the cryptocurrency market, which has lost its relative dynamical autonomy and become significantly tied to the traditional financial instruments. In this work, we present quantitative arguments in support of this statement.

Since the inception of Bitcoin in 2009, the cryptocurrency market has experienced a rapid surge. Although it used to be a niche and traded unofficially in its early years~\cite{pizzaday}, trading takes place now 24/7 on more than 500 exchanges{~\cite{coinmarket}}. The current (October 2022) capitalization of the cryptocurrency market is approximately 1 trillion USD~\cite{coinmarket}, which can be compared to the largest US tech stocks. During these 12 years of Bitcoin history, there were bubbles and crashes~\cite{Gerlach2018,Bellon2022,Zitis2022}. In particular, the foundation of Ethereum in 2015, which allowed for a new application of the blockchain technology in the form of smart contracts, and the subsequent Initial Coin Offer bubble~\cite{Aste2019} in 2018 reshaped the cryptocurrency market and made it appear in the public eye. A recent bubble in 2021, which was related to the adoption of DeFi (Decentralized Finance) and DEX (Decentralized Exchanges) trading~\cite{Maouchi2022}, ended with a peak in November 2021, when the total market capitalization was close to 3 trillion dollars. Although there are more than 10,000 cryptocurrencies~\cite{coinmarket}, Bitcoin and Ethereum are currently the most recognizable, and their share in the capitalization of the entire market changed from over 80\% in early 2021 to 60\% in October 2022~\cite{coinmarket}.

During these 12 years of development, the characteristics of the cryptocurrency market have changed significantly~\cite{watorek2021,Lahmiri2022}. The properties of the cryptocurrency price return time series are now close to those observed in mature financial markets such as Forex~\cite{DrozdzBTC2018}. However, it has long been believed that the cryptocurrency market, which itself is strongly correlated~\cite{Chaos2020,Kaiser2020,Aslanidis2021,entropy2021b,Bae2022,James2022,KAKINAKA2022}, especially during the Covid-19 period~\cite{Arouxet2022,CORBET2022,Crane2022,Kumar2022,watorekfutnet2022}, has dynamics that is separate from the traditional financial markets~\cite{Corbet2018,Ji2018,drozdzfutnet2019,Alana2020,Manavi2020} and that bitcoin can even serve as a hedge or safe haven~\cite{Urquhart2019,Wang2019} with respect to the stock market, Forex or the commodity market. The hedging potential of bitcoin was even compared to gold~\cite{Shahzad2019,Shahzad2019a,bouri2020,Thampanya2020,ZZhu2022}. However, the results of many recent studies have changed this paradigm~\cite{Conlon2020b,Kristoufek2020,Grobys2021,James2021b,Barbu2022,Jareno2022,KAKINAKA2022b,Mandaci2022}. They reported that during the Covid-19 pandemic outburst and the related crash in March 2020~\cite{Baker2020,Yarovaya2022} the cryptocurrency market and, in particular, bitcoin was highly correlated with the falling stock markets~\cite{Conlon2020,entropy2020,Caferra2021,Jiang2021,James2021,Elmelk2022,Wang2022}. Some studies even noted that this connection still occurred in the market recovery phase in the second half of 2020~\cite{entropy2021b,Balcilar2022}.

The studies referenced above brought therefore rather mixed results and have led to uncertainty as to whether cryptocurrencies can be used for hedging the financial investments. This uncertainty opens space for further research on this topic and our study proceeds exactly in this direction. Our aim is to clarify whether the loss of the cryptocurrency market independence was temporary and primarily caused by the pandemic turmoil or it was only a part of a more general trend towards coupling of this market with the traditional financial markets. We intend to determine how strongly the cryptocurrency price changes are associated with the price changes in the traditional financial markets. To achieve that, the detrended multiscale correlation of the two principal cryptocurrencies: bitcoin (BTC) and ethereum (ETH) versus the traditional financial instruments: stock indices, commodities and currency exchange rates are studied based on high frequency data covering the period from January 2020 to October 2022, which is an extension of the period before 2020 that was analyzed in our earlier study~\cite{entropy2020}. The year 2022 is particularly interesting, because since the BTC price peak in November 2021, there is a joint bear market on the US tech stocks and the cryptocurrencies for the first time in the existence of the latter. On the basis of this observation, it is likely that there will be some detectable correlation between both markets. The year 2022 is also unique in the history of cryptocurrencies due to the presence of high inflation in the world against which Bitcoin was intended to protect~\cite{Conlon2021inf,Choi2022,James2022inf,Phochanachan2022}.

Compared to the other articles dealing with correlations between the cryptocurrency market and traditional financial markets, in our research, the main task is to measure these correlations quantitatively on various time scales and for the fluctuations of various size. It can broaden the market practitioners' perspective on the investing and hedging possibilities by incorporating the fluctuation size as an additional dimension that might be considered while making investment decisions.

\section{Data and Methodology}
\label{sect::data.methodology}

\subsection{Data Sources and Preprocessing}

In the present study, a data set of {24} financial time series representing contracts for difference (CFDs) from the Dukascopy trading platform~\cite{Dukascopy} is considered. Unlike many other trading platforms, Dukascopy offers freely the high-frequency recordings of many financial instruments, which is the main reason it has been chosen as the data source. CFDs are characterised by the price movements that are close to the price movements of the original instruments, so we consider them as reliable proxies. Apart from the two highest capitalized cryptocurrencies, BTC and ETH, it includes the most important traditional financial instruments: 12 fiat currencies (Australian dollar---AUD, Canadian dollar---CAD, Swiss franc---CHF, Chinese {yuan}---CNH, euro---EUR, British pound---GBP, Japanese {yen}---JPY, Mexican peso---MXN, Norwegian krone---NOK, New Zealand dollar---NZD, Polish zloty---PLN, and South African rand---ZAR), 4 commodities (WTI crude oil---CL, high grade copper---HG, silver---XAG, and gold---XAU), 4 US stock market indices (Nasdaq 100---NQ100, S\&P500, {Dow Jones Industrial Average}---DJI, and Russell 2000---RUSSEL), German stock index---{DAX 40}---DAX, {and the Japanese stock index---Nikkei 225---NIKKEI}. All these instruments except for the non-US stock indices are expressed in USD (thus there is no USD in the data set) and their quotes cover a period from  1 January 2020 to  28 October 2022. Each week the quotes were recorded over the whole trading hours, i.e., from Sunday 22:00 to Friday 20:15 with a break between 20:15 and 22:00 each trading day (UTC)~\cite{Dukascopy}.

\begin{figure}[H]
\includegraphics[width=0.95\textwidth]{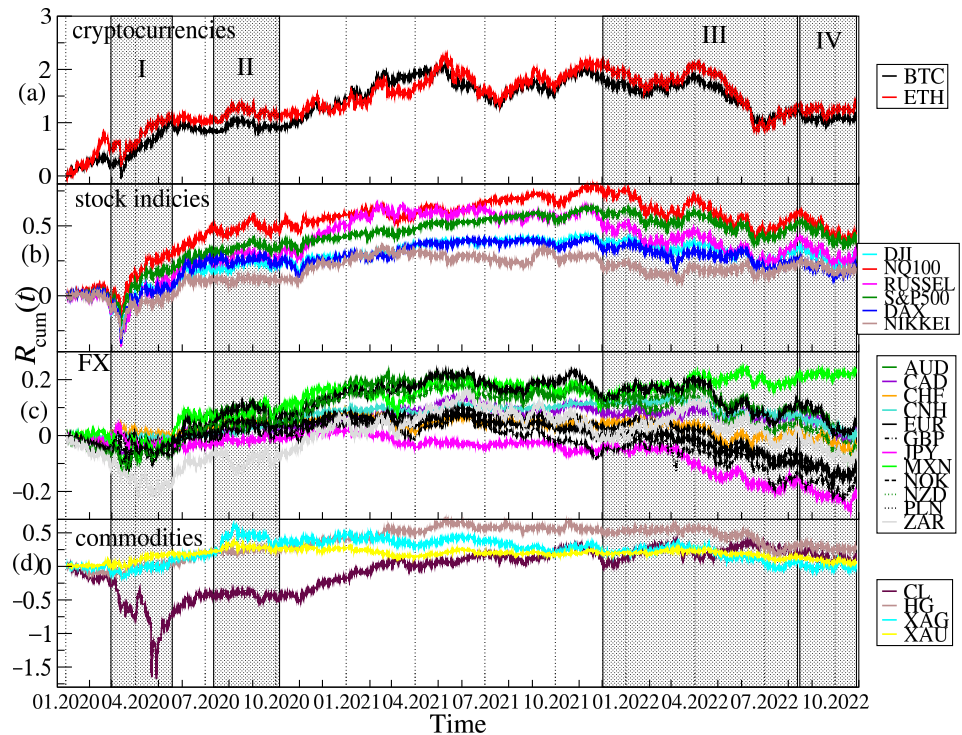}
\caption{Evolution of the cumulative log-returns of the cryptocurrencies $R_{\textrm{cum}}$ (\textbf{a}), the stock market indices (\textbf{b}), the fiat currencies (\textbf{c}), and the commodities (\textbf{d}) over a period from 1 January 2020 to 28 October 2022. Periods for which significant correlations between the cryptocurrencies and the US stock indices are distinguished by grey vertical strips. The most characteristic periods are denoted by Roman numerals: a Covid-19-related crash in March 2020 and a quick bounce in Apr-May 2020 (period I), new all-time highs of NQ100 and S\&P500 and a September 2021 correction (period II), a bear phase in the cryptocurrency and stock markets since November 2021 (period III), and another downward wave of US stock indices after holiday upward correction along with the appreciating USD and inflation fears (period IV).}
\label{fig::price.changes}
\end{figure}

\begin{figure}[H]
\includegraphics[width=0.95\textwidth]{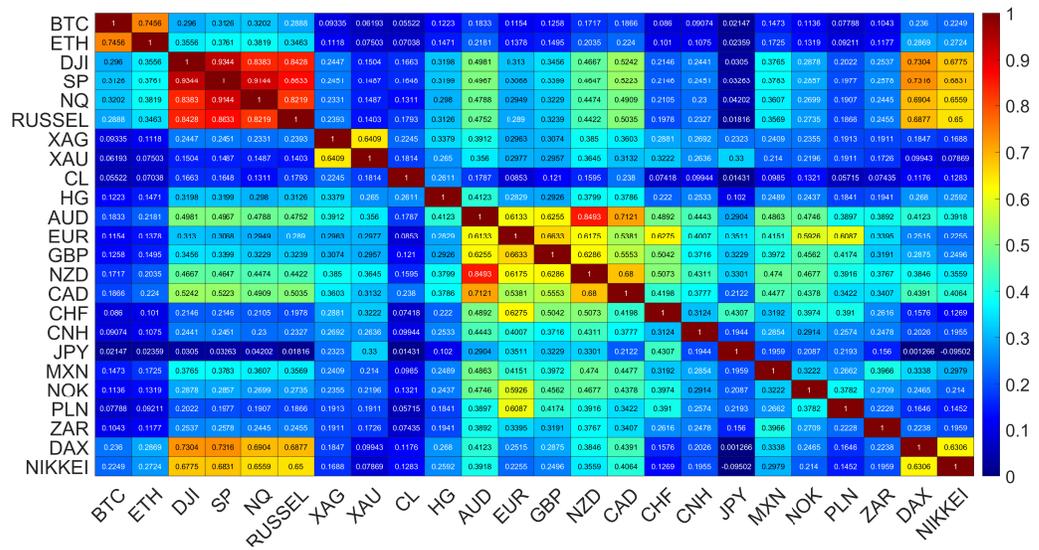}
\caption{Correlation matrix of Pearson coefficients calculated for all possible pairs of the time series considered in this study (January to October 2022). All the values are statistically significant with $p$-value < 0.00001, except DAX vs. JPY, where $p=0.1$.}
\label{fig::correlation.matrix}
\end{figure}

Cumulative log-returns of all the instruments considered are plotted in Figure~\ref{fig::price.changes} against time. The original price changes, sampled every $\Delta t=10s$, were transformed into logarithmic returns: $r(t_m)=\ln P_i(t_{m+1})-\ln P_i(t_m)$, where $P_i(t_m)$ is a price quote recorded at time $t_m$ ($m=1,\dots ,T$) and $i$ stands for a particular financial instrument. We use this particular time interval $\Delta t$, because such a data set was available from the source. However, it is satisfactory because it allows us to avoid excessive null returns, which lower reliability of the detrended analysis (see below). In order to obtain the indicative relationships among all the time series, the Pearson correlation coefficient $C$~\cite{Pearson1895} was calculated for the log-returns $r(t_m)$ from January to October 2022, when the joint bear market mentioned above was observed. A correlation matrix obtained for {24} financial instruments is shown in Figure~\ref{fig::correlation.matrix}. While the Pearson coefficient is one of the most widely applied measures of time series dependencies (and this is why we also exploited it in our study), the results obtained with it have to be taken with some reserve in our context. This is because the statistical tests that we carried out, i.e., the Jacque-Bera test for normality and the ARCH test for no heteroskedasticity, both rejected the respective null hypotheses with high confidence ($p$-value < 0.00001), which means that the data under study was both heavy-tailed and heteroskedastic. Obviously, such a result is not surprising, because fat tails of the return distributions and volatility clustering are well-known effects observed in the financial time series~\cite{Gopikrishnan1999,Cont2001,Gabaix2003}. Nevertheless, the very long time series that were analysed here and the statistical significance of the obtained results convinced us that the Pearson coefficient can still serve as an effective measure of the time series correlations even in such circumstances. Taking all this into account, a standard naming convention: small ($0.1 \le C < 0.3$), medium ($0.3 \le C < 0.$), and large ($0.5 \le C \le 1.0$) correlation was used to describe the coefficient values. The strongest cross-correlations ({large}, $C>0.6$) can be seen {for the stock indices}, for BTC and ETH, for AUD, NZD and CAD, for XAU and XAG, and for EUR and GBP. If we consider the cross-correlations between BTC and the traditional instruments, the strongest ones can be seen for NQ100 and S\&P500 ({medium}, $C\approx0.32$), DJI and RUSSEL ({medium}, $C\approx0.29$), DAX ({small}, $C\approx0.24$) and {NIKKEI, (small $C\approx0.23$)}. The Pearson coefficient above 0.1 {(small)}, is observed for BTC on the one side and HG, GBP and EUR, as well as the so-called commodity currencies: AUD, CAD, NZD, MXN, NOK, and ZAR, on the other side. The cross-correlations between ETH and the other instruments are even higher: $C\approx0.38$ {(medium)} for SP500 and NQ100, $C\approx0.35$ {(medium)} for DJI and RUSSEL, $C\approx0.29$ {(medium)} for DAX, and $C\approx0.27$ {(small)} for NIKKEI. The same is true for the cross-correlations between ETH and the commodity currencies: $C\approx0.22$ {(small)} for AUD, CAD, NZD, $C\approx0.17$ {(small)} for MXN, $C\approx0.13$ {(small)} for NOK, and $C\approx0.12$ {(small)} for ZAR. Among the commodities analyzed here, ETH is the most correlated with HG ($C\approx0.15$, {small}). {The statistical significance of the  coefficient values presented in Figure~\ref{fig::correlation.matrix} has been checked by calculating the range: $\bar{C} \pm \sigma_C$, where $\bar{C}$ denotes mean and $\sigma_C$ denotes standard deviation of $C$, from 100 independent realisations of the shuffled time series. The statistically insignificant correlation region is very close to 0 (the third decimal place), all the presented values, except DAX vs. JPY, are thus significant.}

\unskip

\subsection{The $q$-Dependent Detrended Correlation Coefficient}

Since the Pearson correlation coefficient as a measure has its drawbacks in the case of heavy tails, heteroskedasticity, and multi-scale nonstationarity (which are observed in the cryptocurrency market~\cite{watorek2021}) the cross-correlations will henceforth be determined using an alternative, better tailored method: the $q$-dependent detrended cross-correlation coefficient $\rho_q(s)$~\cite{kwapien2015}. The detrended fluctuation analysis (DFA), which forms a basis for defining $\rho_q(s)$, was developed with the intention to allow for detecting the long-range power-law auto- and cross-correlations that produce trends on different time horizons~\cite{PengCK-1994a}. Therefore, unlike more traditional methods of trend removal, both DFA and its derivative measures like the coefficient $\rho_q(s)$ can successfully deal with nonstationarity on all scales~\cite{Jiang_2019}. $\rho_q(s)$ enables, thus, considering the cross-correlation strength on different time scales and, if used in parallel with the multiscale DFA itself, is able to detect scale-free correlations. Moreover, owing to the parameter $q$, the correlation analysis can be focused on a specific range of the fluctuation amplitudes.

The steps to calculate $\rho_q(s)$ are as follows. Two possibly nonstationary time series $\{x(i)\}_{i=1,\dots ,T}$ and $\{y(i)\}_{i=1,\dots ,T}$ of length $T$ are divided into $M_s$ boxes of length $s$ starting from its opposite ends and integrated. In each box, the polynomial trend is removed:
\begin{align}
X_{\nu}(s,i) = \sum_{j=1}^i x(\nu s + j) - P^{(m)}_{X,s,\nu}(i), 
Y_{\nu}(s,i) = \sum_{j=1}^i y(\nu s + j) - P^{(m)}_{Y,s,\nu}(i),
\end{align}
where the polynomials $P^{(m)}$ of order $m$ are applied. In this study $m=2$ has been selected, which performs well for the financial time series~\cite{oswiecimka2013}. After this step $2 M_s$ boxes are obtained in total with detrended signals. The next step is to calculate the variance and covariance for each of the boxes $\nu$:
\begin{align}
f^2_{\rm ZZ} (s,\nu) = \frac{1}{s}\sum_{i=1}^s (Z_{\nu}(s,i) )^2, \\
f^2_{\rm XY} (s,\nu) = \frac{1}{s}\sum_{i=1}^s X_{\nu}(s,i)\times Y_{\nu}(s,i),
\end{align}
where $Z$ means $X$ or $Y$. These quantities are used to calculate a family of the fluctuation functions of order $q$:
\begin{align}
F^{(q)}_{\rm ZZ} (s) = {1 \over 2 M_s} \sum_{\nu=0}^{2 M_s-1} \left[ f^2_{\rm ZZ} (s,\nu)\right]^{q/2}
\label{eq::fq.zz} \\
F^{(q)}_{\rm XY} (s) = {1 \over 2 M_s} \sum_{\nu=0}^{2 M_s-1} \textrm{sign} \left[ f^2_{\rm XY}(s,\nu)\right] |f^2_{\rm XY} (s,\nu)|^{q/2},
\label{eq::fq.xy}
\end{align}
where a sign function $\textrm{sign} \left[ f^2_{\rm XY}(s,\nu)\right]$ is preserved in order to secure consistency of results for different $q$s.

The formula for the $q$-dependent detrended correlation coefficient is given as follows:
\begin{equation}
\rho_q^{\rm XY}(s) = {F^{(q)}_{\rm XY}(s) \over \sqrt{F^{(q)}_{\rm XX}(s) F^{(q)}_{\rm YY}(s)}}.
\label{eq::rhoq}
\end{equation}

For $q=2$ the above definition can be viewed as a detrended counterpart of the Pearson cross-correlation coefficient $C$~\cite{zebende2011}. The parameter $q$ plays the role of a filter suppressing $q<2$ or amplifying ($q>2$) the fluctuation variance/covariance calculated in the boxes $\nu$ (see Eqs.~(\ref{eq::fq.zz}) and (\ref{eq::fq.xy})). For $q<2$ boxes with small fluctuations contribute more to $\rho_q(s)$, while for $q>2$ the boxes with large fluctuations contribute more. Therefore, by using this measure, it is possible to distinguish the fluctuation size range that is a source of the observed correlations. In the numerical calculations below, we used our own software in which we implemented the algorithm described above.

\section{Results and Discussion}
\label{sect::results.discussion}

\begin{figure}[H]
\includegraphics[width=1\textwidth]{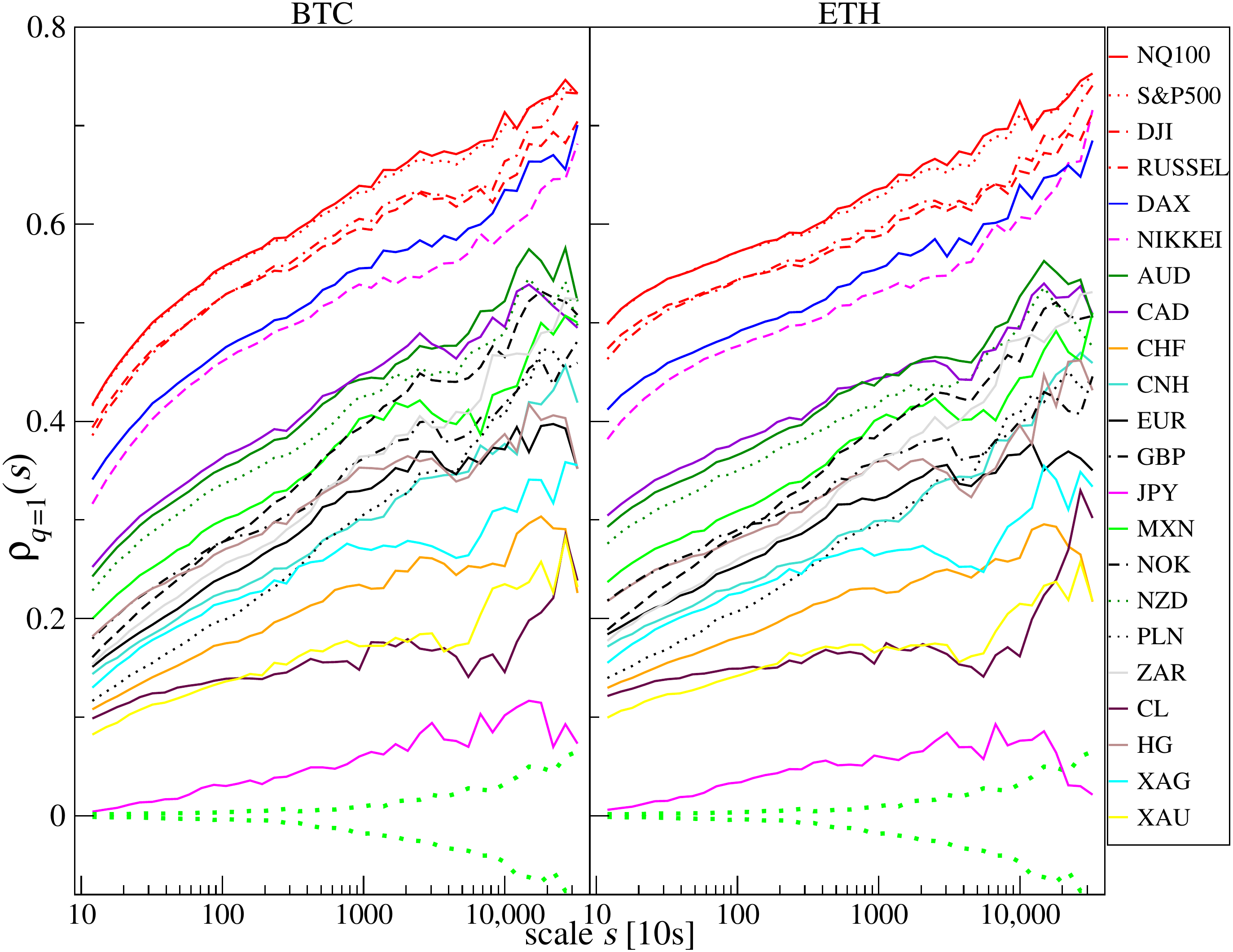}
\caption{The $q$-dependent detrended cross-correlation coefficient $\rho_q(s)$ between BTC/USD (right) and ETH/USD (left) versus selected traditional financial instruments for {$q=1$}, which does not favor any specific amplitude range. $\rho_q(s)$ for a range of time scales from $s=12$ (2 min) to $s=32,000$ ($\sim$4 trading days) is presented based on data from January to October 2022. The statistically insignificant correlation region (dotted green line) is given as $\pm$ standard deviation of $\rho_q(s)$ calculated from 100~independent realizations of the shuffled time series.}
\label{fig::rhoq1}
\end{figure}

\begin{figure}[H]
\includegraphics[width=1\textwidth]{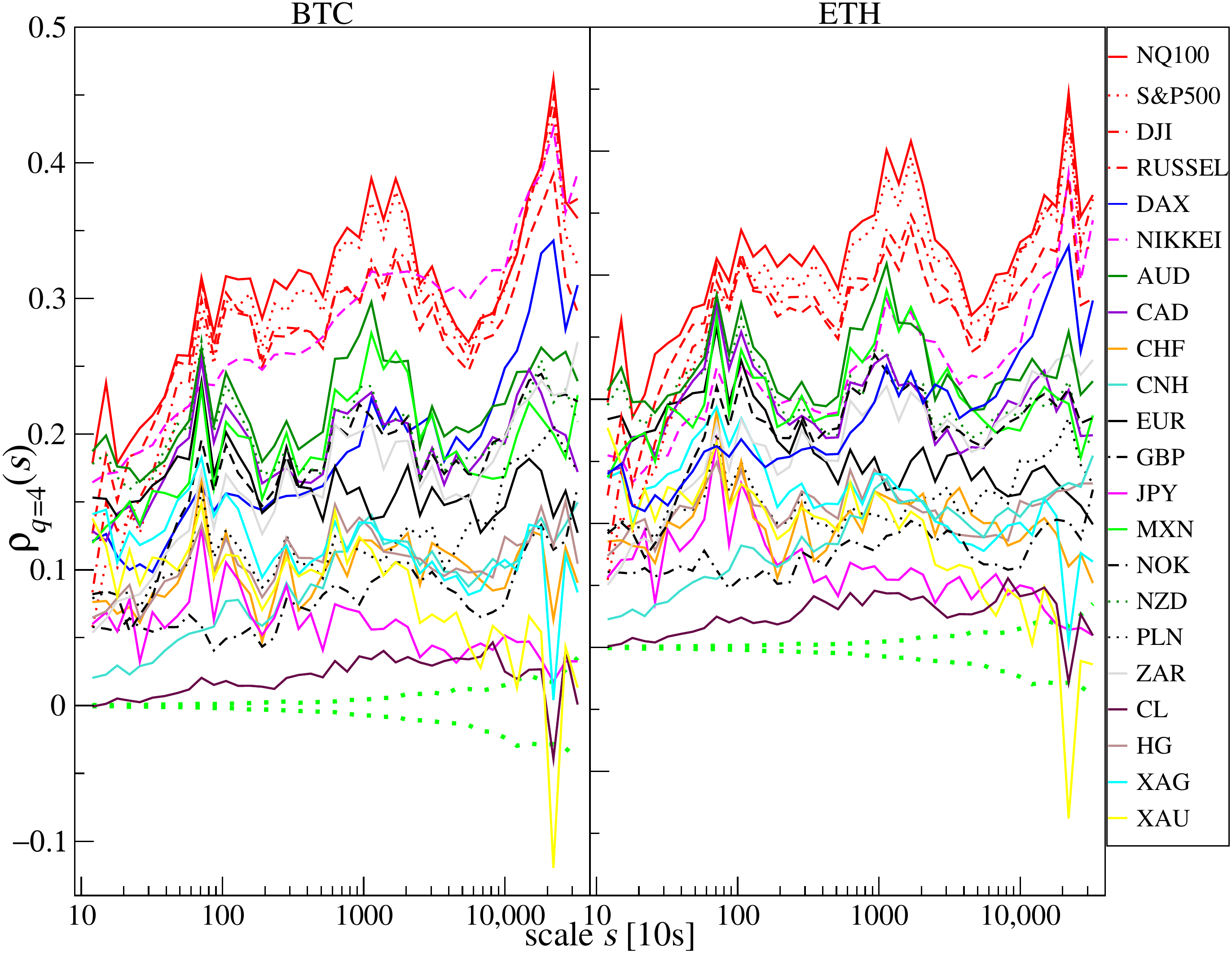}
\caption{The $q$-dependent detrended cross-correlation coefficient $\rho_q(s)$ between BTC/USD (right) and ETH/USD (left) versus selected traditional financial instruments for {$q=4$}, which amplifies the large return contributions. $\rho_q(s)$ for a range of time scales from $s=12$ (2 min) to $s=32,000$ ($\sim$4 trading days) is presented based on data from January to October 2022. The statistically insignificant correlation region (dotted green line) is given as $\pm$ standard deviation of $\rho_q(s)$ calculated from 100~independent realizations of the shuffled time series.}
\label{fig::rhoq4}
\end{figure}

The aforementioned ability of $\rho_q(s)$ to quantify cross-correlation for various time scales ($s$-dependence) and fluctuation size ($q$-dependence) is documented in {Figures~\ref{fig::rhoq1}~and~\ref{fig::rhoq4}}, where the values of $\rho_q(s)$ calculated for BTC and ETH versus the traditional instruments (the same as Figure~\ref{fig::price.changes}) calculated for the log-returns $r(t_m)$ from January to October 2022 is shown. One can immediately notice two properties: (i) the correlation strength increases with scale $s$ for most financial instruments, and (2) the correlation strength is lower for $q=4$ (i.e., for large fluctuations {Figure~\ref{fig::rhoq4})}. These properties, observed here for BTC and ETH versus the other instruments, are typical for the financial markets in general~\cite{gebarowski2019,Watorek2019}.

As in the case of the Pearson coefficient, the strongest cross-correlations measured by $\rho_q(s)$ for $q=1$ {(Figure~\ref{fig::rhoq1})} are BTC and ETH versus the stock indices NQ100 and S\&P500. It is different for DJI, RUSSEL, DAX, {and NIKKEI}, which are less cross-correlated with the cryptocurrencies. What is interesting is that these correlations were stronger for ETH than for BTC, particularly on short time scales. For the shortest scale considered \mbox{($s=12=2$ min),} they started from $\rho_q(s)\approx0.5$ in the case of ETH vs. NQ100 and S\&P500 and from $\rho_q(s)\approx0.4$ in the case of BTC vs. NQ100 and S\&P500. For the longest scale considered ($s=32,000\approx4$ trading days), the coefficient $\rho_q(s)\approx 0.75$ for BTC and ETH vs. NQ100 and S\&P500. The lowest correlations and the weakest scale dependence are observed for JPY, where $\rho_q(s)\approx0$. XAU and CL are slightly more correlated: $\rho_q(s)\approx0.1$ and 0.2 for the longest scale $s$. Above them are XAG and CHF for which the correlations increase with $s$ from 0.1 to 0.3. The cross-correlations for remaining fiat currencies and HG start from $\rho_q(s)\approx0.15\div0.25$ for $s=12$ and end at $\rho_q(s)\approx0.35\div0.55$ for $s=32,000$. If we focus on the large fluctuations and apply $q=4$ {(Figure~\ref{fig::rhoq4})}, the cross-correlation levels are lower and approximately the same for BTC and ETH. Again, the most significant correlations are observed for the BTC and ETH vs. the US stock indices, but {the correlations between BTC and NIKKEI are higher by $\sim0.05$ than for ETH and NIKKEI. In the range of scales $4000 \le s \le 10,000$ the correlations between BTC and NIKKEI are the strongest. Unlike for $q=1$, the} cross-correlations {for} BTC and ETH vs. DAX are on the same level as vs. AUD, CAD, MXN, NZD, and NOK. Only for $s\approx20,000$ the negative values of $\rho_q(s)$ can be found for BTC and ETH versus XAU. The statistical significance of $\rho_q(s)$ in each case was determined by calculating the standard deviation of $\rho_q(s)$ for 100 independent realizations of shuffled time series. This quantity is plotted in Figures~\ref{fig::rhoq1}~and~\ref{fig::rhoq4} by green dotted lines. It shows that the detrended cross-correlations are significant for all the instruments in the case of $q=1$, except for the longest considered scales for JPY, while in the case of $q=4$, the results {for CL and XAU lack statistical significance for the longest considered scales.}

\begin{figure}[H]
\centering
\includegraphics[width=0.99\textwidth]{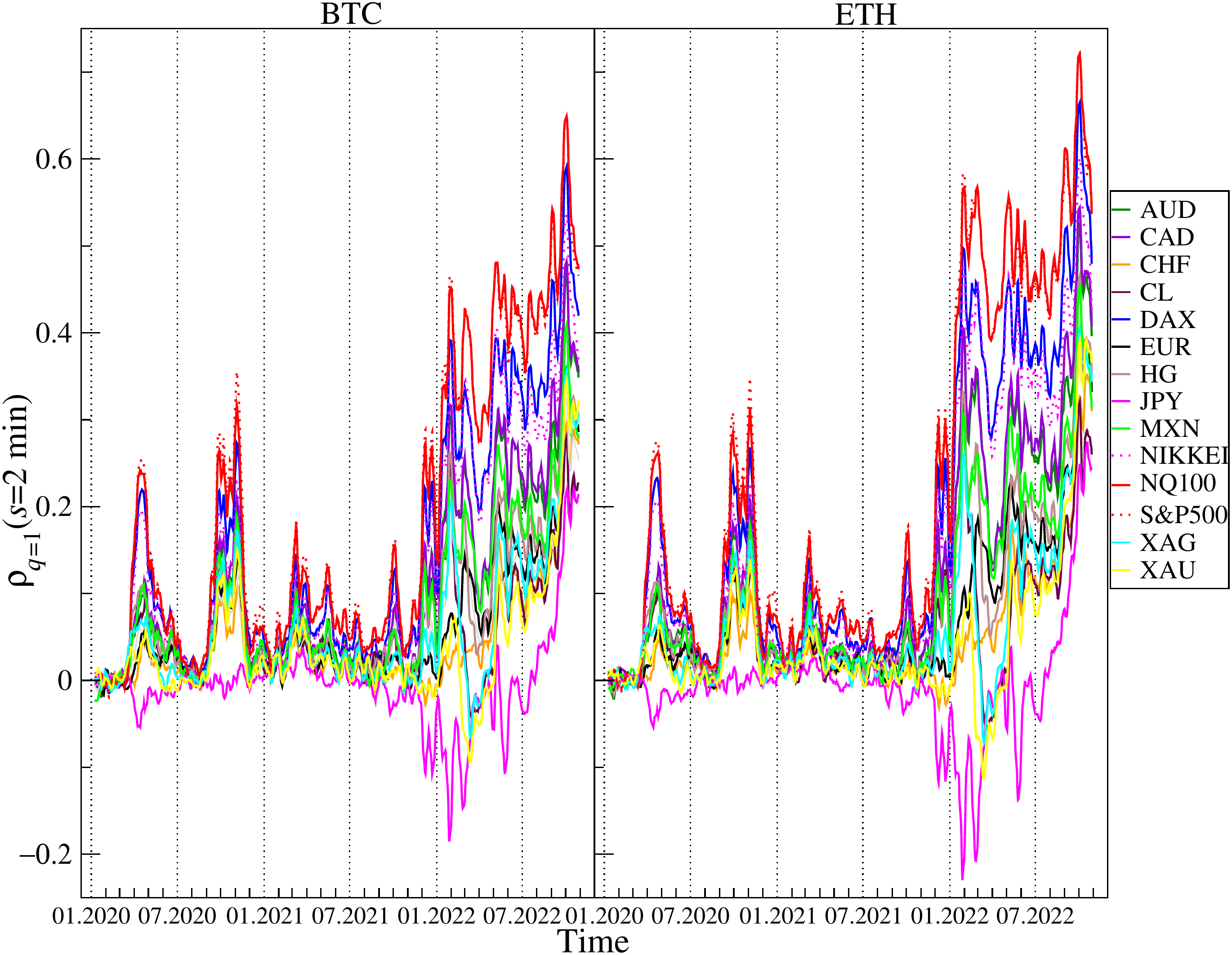}
\caption{Evolution of the $q$-dependent detrended cross-correlation coefficient $\rho_q(s)$ with $q=1$ and {$s=2$} min calculated in a 5-day rolling window with a 1-day step between 1 January 2020 and 28 October 2022 for the price returns of BTC/USD (left) and ETH/USD (right) versus the selected traditional assets: AUD, CAD, CHF, CL, DAX, EUR, HG, JPY, MXN, NIKKEI, NQ100, S\&P500, XAG, and XAU. {The statistically insignificant correlations are in the region $\rho_q(s)=0\pm0.001$.}}
\label{fig::evolution.q1s12}
\end{figure}

\begin{figure}[H]
\centering
\includegraphics[width=1\textwidth]{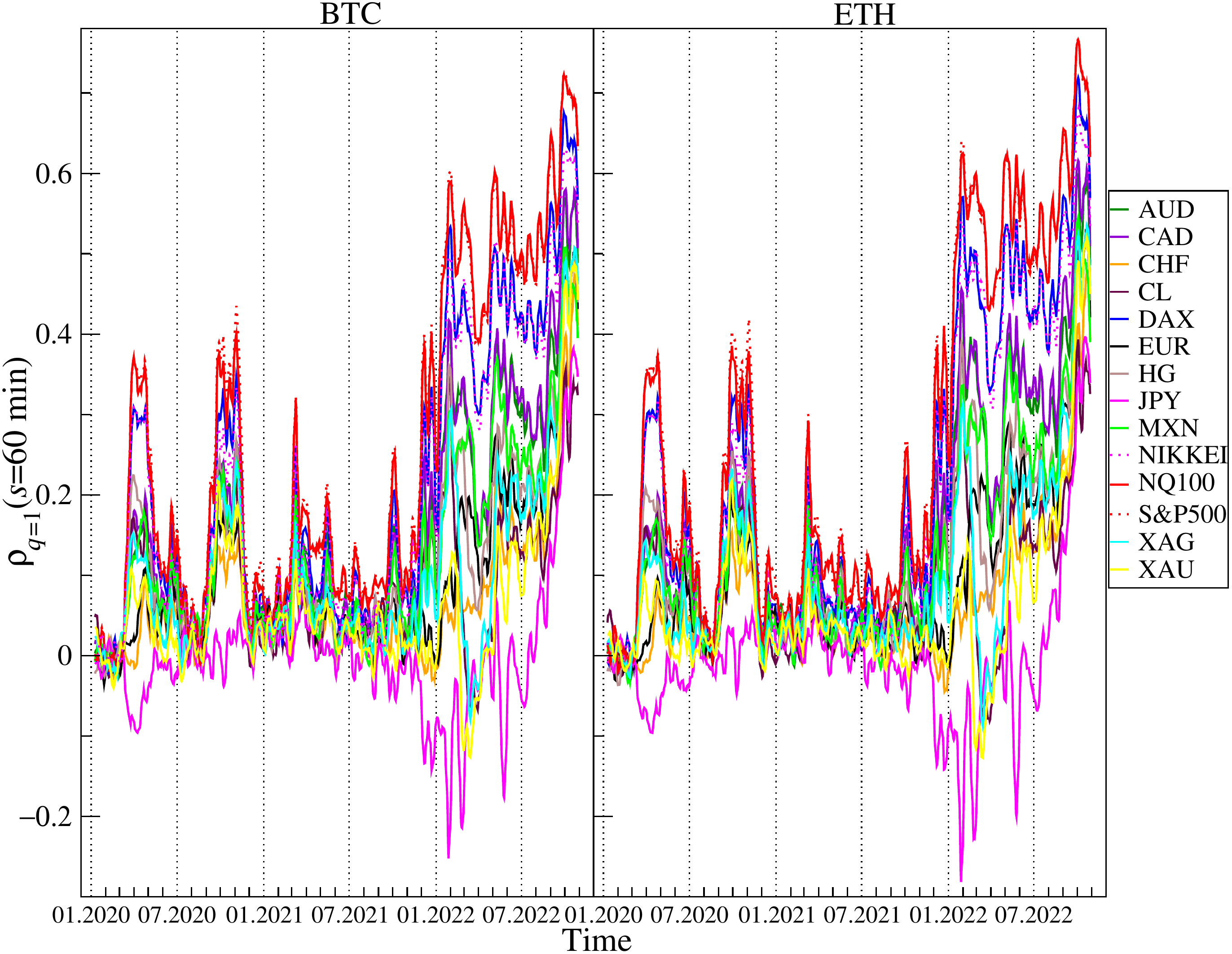}
\caption{Evolution of the $q$-dependent detrended cross-correlation coefficient $\rho_q(s)$ with $q=1$ and {$s=60$} min calculated in a 5-day rolling window with a 1-day step between 1 January 2020 and 28 October 2022 for the price returns of BTC/USD (left) and ETH/USD (right) versus the selected traditional assets: AUD, CAD, CHF, CL, DAX, EUR, HG, JPY, MXN, NIKKEI, NQ100, S\&P500, XAG, and XAU. {The statistically insignificant correlations are in the region $\rho_q(s)=0\pm0.01$.}}
\label{fig::evolution.q1s60}
\end{figure}

\begin{figure}[H]
\includegraphics[width=1\textwidth]{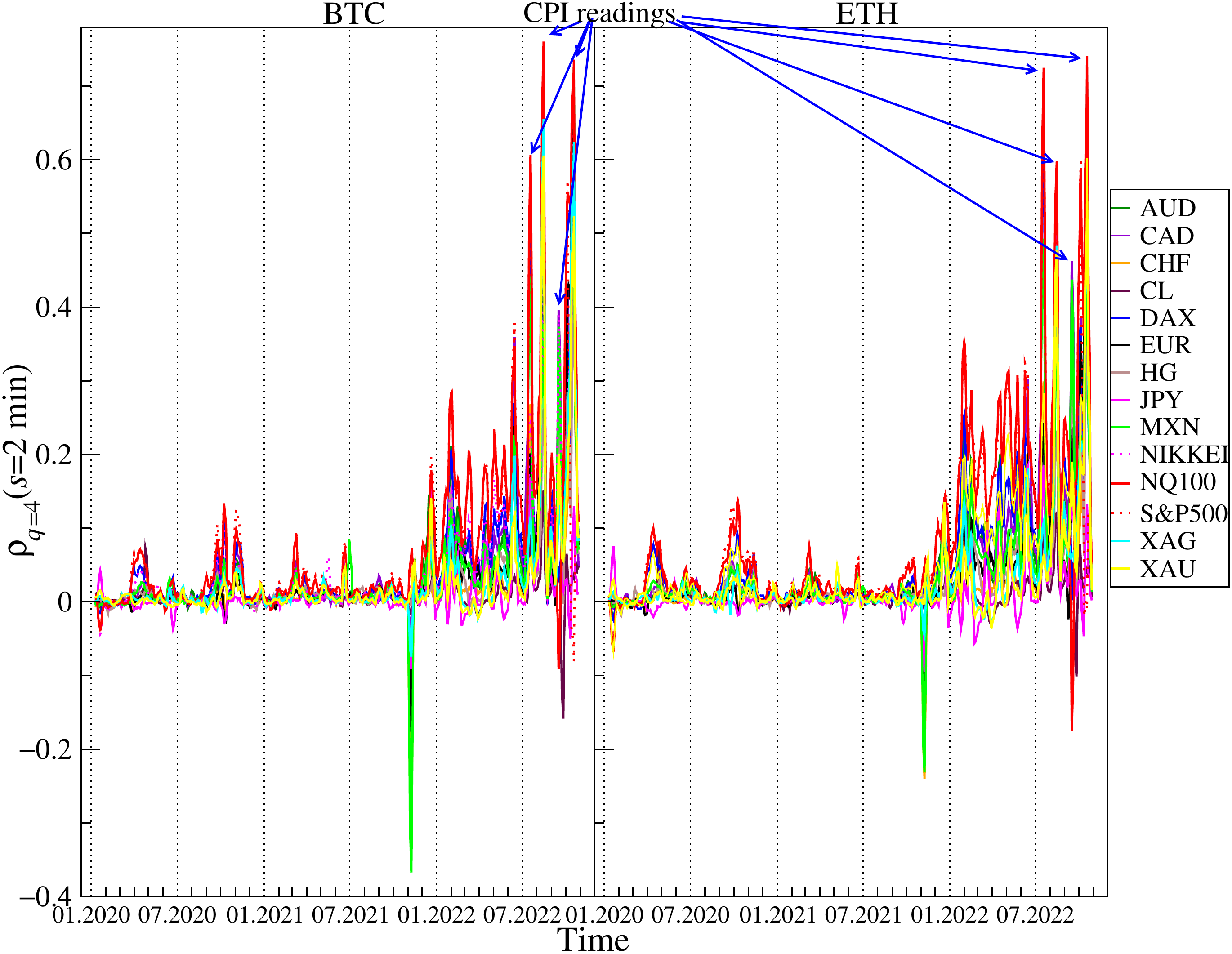}
\caption{Evolution of the $q$-dependent detrended cross-correlation coefficient $\rho_q(s)$ with $q=4$ and {$s = 2$} min calculated in a 5-day rolling window with a 1-day step between 1 January 2020 and 28 October 2022 for the price returns of BTC/USD (left) and ETH/USD (right) versus the selected traditional assets: AUD, CAD, CHF, CL, DAX, EUR, HG, JPY, MXN, NIKKEI, NQ100, S\&P500, XAG, and XAU. Higher levels of cross-correlations, associated with the {Consumer Price Index (CPI)} readings, are marked (see the event description in Figure~\ref{fig::price.changesCPI}). {The statistically insignificant correlations are in the region $\rho_q(s)=0\pm0.001$.}}
\label{fig::evolution.q4s12}
\end{figure}

\begin{figure}[H]
\includegraphics[width=1\textwidth]{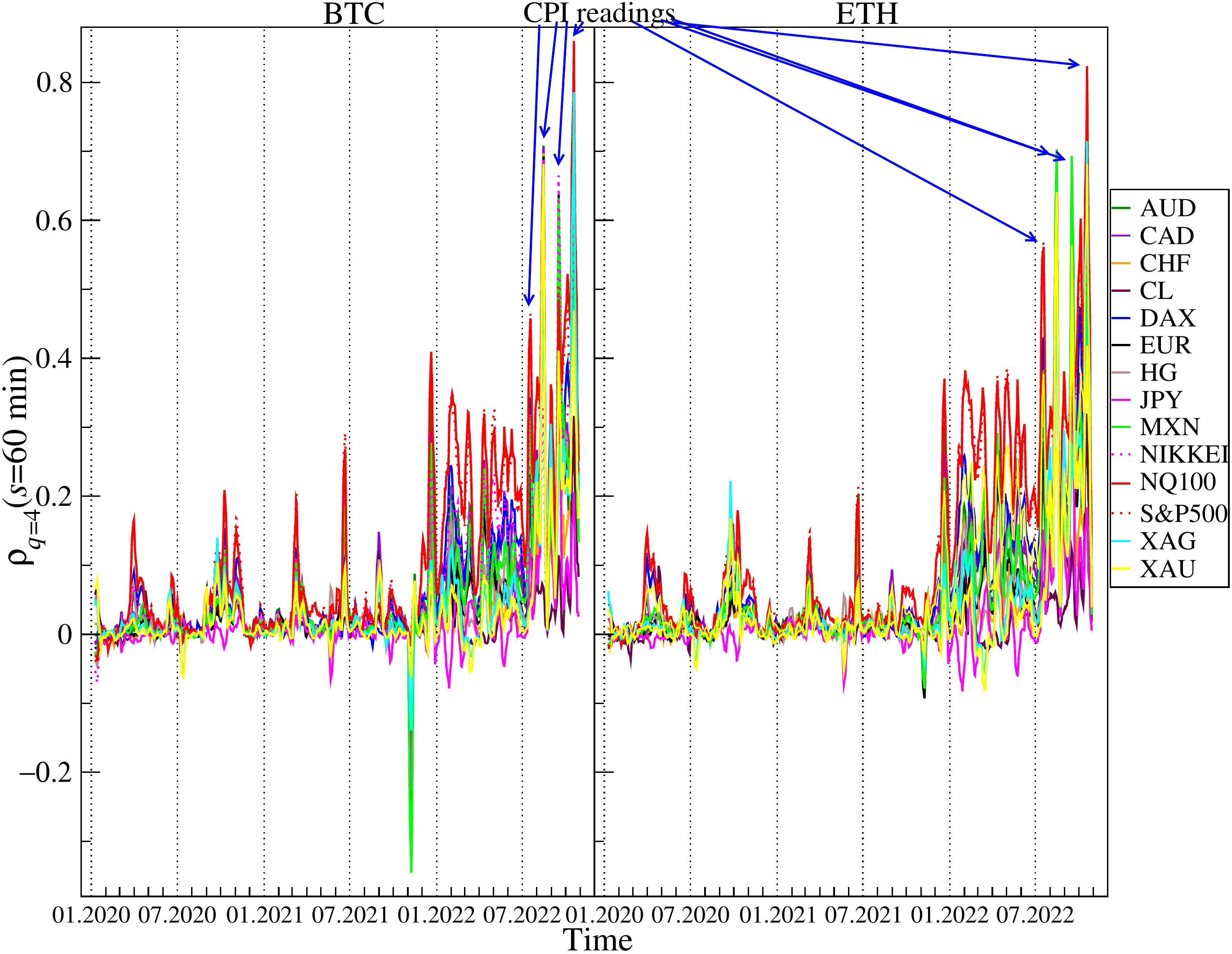}
\caption{Evolution of the $q$-dependent detrended cross-correlation coefficient $\rho_q(s)$ with $q=4$ and {$s = 60$} min calculated in a 5-day rolling window with a 1-day step between 1 January 2020 and 28 October 2022 for the price returns of BTC/USD (left) and ETH/USD (right) versus the selected traditional assets: AUD, CAD, CHF, CL, DAX, EUR, HG, JPY, MXN, NIKKEI, NQ100, S\&P500, XAG, and XAU. Higher levels of cross-correlations, associated with the {Consumer Price Index (CPI)} readings, are marked (see the event description in Figure~\ref{fig::price.changesCPI}. The statistically insignificant correlations are in the region $\rho_q(s)=0\pm0.01$.}
\label{fig::evolution.q4s60}
\end{figure}

Now, a time-dependent analysis of the cross-correlations measured by $\rho_q(s)$ for BTC and ETH versus the traditional financial instruments: AUD, CAD, CHF, CL, DAX, EUR, HG, JPY, MXN, NIKKEI, NQ100, S\&P500, XAG and XAU will be presented. A 5-day rolling window with a 1-day step was applied on two time scales: $s=12$ (2 min) and $s=360$ (60~min) in order to calculate $\rho_q(s)$. A window of this length corresponds to a trading week. Figures~\ref{fig::evolution.q1s12}~and~\ref{fig::evolution.q1s60} shows the results obtained for $q=1$ and {Figures~\ref{fig::evolution.q4s12}~and~\ref{fig::evolution.q4s60}} shows the results obtained for $q=4$. The results for some assets presented in Figures~\ref{fig::rhoq1}~and~\ref{fig::rhoq4} are omitted here because they are similar to the ones already shown. Our previous study~\cite{entropy2020} reported that before 2020 the cross-correlations for BTC and ETH versus the traditional instruments were close to 0. In this study, the period starting in 2020 is considered, thus. During these 2.5 unstable years, several important events that affected price changes in the financial markets could be observed.

The first event was the outburst of the Covid-19 pandemic that caused a crash in March 2020 on almost all the financial instruments expressed in USD. Only JPY and CHF gained in early March 2020, but later on they also lost value against the US dollar. This price behavior during period I (see Figure~\ref{fig::price.changes}) resulted in the appearance of a significant positive cross-correlation for BTC and ETH versus the risky assets such as the stock indices, CL, HG, and the commodity currencies (AUD, NZD, CAD, MXN, NOK), which can be seen in {Figures~\ref{fig::evolution.q1s12}~and~\ref{fig::evolution.q1s60}}. The largest values of $\rho_q(s)$ for BTC and ETH versus the stock indices are observed. In the case of $q=1$ and $s=2$ min, $\rho_q(s)\approx0.2$ and in the case of $q=1$ and $s=60$ min, $\rho_q(s)\approx0.4$. Such strong cross-correlations observed during the general meltdown are not that surprising, but still the joint behavior of the cryptocurrencies and, particularly, the stock indices is noteworthy because it has changed the view that the cryptocurrency market is independent. What is more interesting is the appearance of the even stronger positive cross-correlations for BTC and ETH versus almost all the other instruments except for JPY in the second half of 2020. The strongest cross-correlations are observed again for the stock indices, but very close were also those for CL, HG, XAG, XAU, and the commodity currencies. The highest values, $\rho_q(s)>0.5$ for $q=1$ and $s=60$ min, were observed at the turn of September and October 2020 after the stock and cryptocurrency markets peaked and turned down at the beginning of September 2020. The third period of the significant cross-correlations for BTC and ETH versus the other instruments starts at the beginning of December 2021 after the November 2021's all-time highs on both the cryptocurrency and the US stock markets occurred. $\rho_q(s)$ grew above 0.5 for $q=1$ and $s=2$ min and above 0.6 for $q=1$ and $s=60$ min in January 2022, when both markets experienced strong declines. BTC and ETH dropped 50\% from their peak price down to 33,000 USD and 2300 USD, respectively, S\&P500 dropped 8\% down to 4,200 USD and NQ100 dropped 18\% to 13,700 USD that were their local lows on January 22, 2022. At that time, there were also significant negative cross-correlations for BTC and ETH versus JPY, which is typically considered as a safe currency during the market meltdowns. After local peak of cross-correlations at the beginning of May 2022, when S\&P500, NQ100, BTC, and ETH broke into new lows below 4150, 13,000, 35,000, and 2200 levels, respectively, the cross-correlations for BTC and ETH vs. the remaining instruments were significant at approximately the same levels until mid-August 2022, when the holiday upward correction in the US stock indices ended. From that moment on, one can distinguish period IV, when another downward wave of US exchange indices took place, which lasted until mid-October. This was accompanied by a strengthening of the USD, and the EUR/USD exchange rate fell below 1. During this period, the cross-correlation of BTC and ETH with all instruments denominated in USD has started to increase. They were even significantly positive in the case of the least correlated JPY at a level above 0.2 for $s=2$ min and 0.4 for $s=$ 60 min. The cross-correlations peaked in the last week of September, when for NQ100 and S\&P500 they first exceeded the level of 0.6 and in the case of $s = 60$ min, they were close to 0.8. They were again slightly higher for ETH.

If large returns are considered ($q=4$, {Figures~\ref{fig::evolution.q4s12} and \ref{fig::evolution.q4s60})} the detrended cross-correlations for $s=2$ min {remain close to 0} and are statistically insignificant {for most of the considered instruments} until November 2021, when $\rho_q(s)$ for BTC and ETH versus most currencies, especially MXN, CHF, and, to a lesser extent, for AUD, NZD, EUR and CNH turn negative for short periods of time. As in the case of $q=1$, the cross-correlations versus the US indices became significantly positive starting from December 2021. What is most interesting is that from July 2022, the cross-correlation levels in some weekly windows exceed those obtained for $q = 1$. For $s = 60$ min they are even higher than 0.8 in the case of NQ100. There are also high correlations of BTC and ETH vs. precious metals: gold and silver. Unlike average fluctuations ($q=1$), here BTC is slightly more strongly correlated with traditional financial instruments. 

\begin{figure}[H]
\centering
\includegraphics[width=1\textwidth]{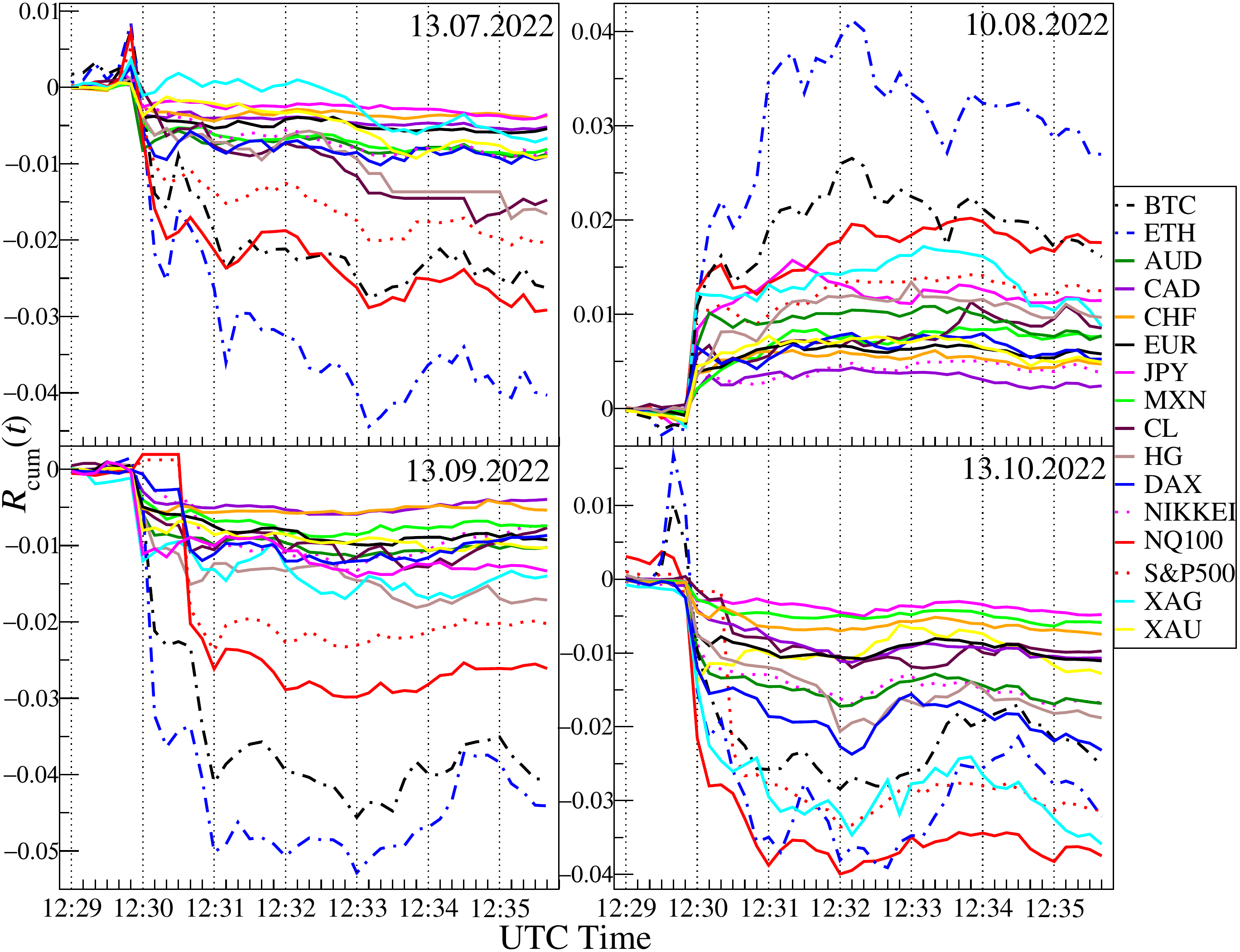}
\caption{Evolution of the cumulative logarithmic returns $R_{\textrm{cum}}$ of selected financial instruments: BTC, ETH, AUD, CAD, CHF, CL, DAX, EUR, HG, JPY, MXN, NIKKEI, NQ100, S\&P500, XAG, and XAU on specific dates, around the publication time of the {Consumer Price Index} (CPI) report in the US.}
\label{fig::price.changesCPI}
\end{figure}

After careful checking of the exact start and end dates of the sliding window with increased correlations for $q = 4$ and the time of large fluctuations, it turned out that the correlation of BTC and ETH with traditional financial instruments is (directly or indirectly via other markets) influenced by the CPI inflation data published every month at 12:30 UTC. Cumulative price changes in days during the CPI publication date 13 July 2022, 10 August 2022, 13 September 2022,  10 October 2022, from 12:29 to 12:35 are presented in the Figure~\ref{fig::price.changesCPI}. It can be clearly seen that in all four cases US tech stocks and cryptocurrencies price changes behave in the same way just after 12:30 UTC. It happened regardless of whether the surprise with the CPI data was positive or negative. In three cases, inflation data was higher than expected and surprised markets negatively, leading to declines. This is especially well visible in the case of  10 October 2022, when apart from US indices, XAG also follows the same trajectory. In the roling windows containing this day, the correlations were the strongest: 0.6 for S\&P500, 0.79 for XAG and 0.86 for NQ100 vs. BTC and 0.6 for S\&P500, 0.72 for XAG and 0.82 for NQ100 vs. ETH for $s=60$ min. In one case, 10 August 2022, the inflation was lower than expectations, which resulted in an increase in all instruments. This price behavior means that cryptocurrencies have started to respond to readings from the economy, just like traditional financial instruments. Despite the fact that our analysis of the cross-correlations was carried out by means of the measures, which were unable to detect the direction of influence, it seems natural to infer that these were the economical data releases that had direct or indirect impact on the cryptocurrency market rather than the opposite. That is why we concluded about the direction above.

\section{Conclusions}
\label{sect::conclusions}

Based on the multiscale cross-correlation analysis performed for the data covering almost the last three years, it can be concluded that the cryptocurrency market dynamics is substantially tied to the traditional financial markets. Consistently, the most liquid cryptocurrencies, BTC and ETH, cannot serve as a hedge or safe haven for the stock market investments, not only during the turbulent periods like the Covid-19 panic, where this effect is particularly strong, but also during the recent bear market period on tech stocks, which has been accompanied by the parallel bear market on cryptocurrencies. Many observations show that the Covid-19 pandemic may have changed the paradigm that the cryptocurrency market is a largely autonomous market. The recent market developments and the strong US dollar have additionally increased the strength of the cross-correlations between BTC and ETH on the one side and the US tech stocks on the other side. These observations support some earlier findings on the same subject (e.g.,~\cite{Zhang2021,Kumar2022}). In contrast, as the cryptocurrency market was weakly correlated with other markets during 2021, our results cannot support directly a recent hypothesis that the quantitative easing could actually be responsible for these correlations~\cite{Kumar2022}. The existence of links between the global economy and the cryptocurrency market are further strengthened by the reaction of the price changes of BTC and ETH to economic data, such as CPI inflation, in a similar way to traditional financial instruments. These results are able to remove or, at least, to suppress the uncertainty that recent literature on this topic has brought to the cryptocurrency investors. Now it is more clear that the cryptocurrencies can no longer serve as a convenient hedging target for the investors whose purpose is to diversify the risk.

Our study brings a strong indication that the cryptocurrency market has finally become a connected part of the global financial markets after 12 years of the maturation process. Whether such a direction of this market evolution remains in agreement with the early vision of the cryptocurrency creators can be disputed, however. We also face a related question: does the fact that we have got ``just another part of the global financial market'' deserve devoting so huge amounts of energy to it? Sooner or later this question must be addressed by the policy makers. Nevertheless, what becomes evident now is that it allows the market participants to broaden the spectrum of their investment possibilities.

Among the limitations that might have influenced our study and, subsequently, our conclusions, we have to mention that only two principal cryptocurrencies were studied. Although they are the most influential, the most frequently traded, and widely discussed cryptocurrencies, they by no means define the entire market. It is possible that an analysis that included some less important cryptocurrencies would bring different outcomes. This is especially likely for the marginal cryptocurrencies without any thinkable ``fundamental'' value, whose dynamics is driven predominantly by extreme speculation. However, as the cryptocurrency market is looked at by the most investors through the lens of BTC and ETH (as their capitalization indicates), this particular limitation does not seem discouraging to us. Currently, these two assets shape the whole cryptocurrency market and we expect them to continue doing it in the nearest future. Another limitation of our analysis is the particular selection of the traditional financial instruments. Indeed, they constitute only a small fraction of the available ones. We are convinced, though, that they are among the most observed and the most influential ones in the context of the global economy, which fully justifies our choice.

A more general observation that the cryptocurrency market has spontaneously coupled to the technological sector of the stock markets by reacting to some trigger provided by the external data inflow resembles analogous effects of the spontaneous emergence of order among the so-far independent degrees of freedom in the various complex systems. However, as complexity allows for flexible reacting of a system to both the external perturbations and internal processes, such effects of ordering in the financial markets have to be eventually counterbalanced by the opposite processes of disordering. Therefore, the market participants must be aware that the inter-market couplings may not last forever and they can significantly be weakened or even removed completely at some point in future. This is why the in-depth studies of the cross-market dependencies have to remain among the principal directions of the cryptocurrency research. Our future work will also deal with energy consumption of the cryptocurrency market.

\vspace{6pt}

\authorcontributions{Conceptualisation, S.D., J.K. and M.W.; methodology, S.D., J.K. and M.W.; software, M.W.; validation, S.D., J.K. and M.W.; formal analysis, M.W.; investigation, S.D., J.K. and M.W.; resources, M.W.; data curation, M.W.; writing---original draft preparation, M.W.; writing---review and editing, S.D., J.K. and M.W.; visualisation, M.W.; supervision, S.D. All authors have read and agreed to the published version of the manuscript.}

\funding{This research received no external funding.}

\dataavailability{Data available freely from Dukascopy~\cite{Dukascopy}.}

\conflictsofinterest{The authors declare no conflict of interest.}

\begin{adjustwidth}{-\extralength}{0cm}

\reftitle{References}


\end{adjustwidth}
\end{document}